# Preserving The Safety And Confidentiality Of Data Mining Information In Health Care: A literature Review

Robinson Onyemechi Oturugbum

School of Computer Science & Information Systems, Northwest Missouri State University 800 University Drive Maryville, MO 64468 USA.
s542366@nwmissouri.edu
otmrobinson@gmail.com

**ABSTRACTS**

*Daily, massive volume of data are produced due to the internet of things' rapid development, which has now permeated the healthcare industry. Recent advances in data mining have spawned a new field of a study dubbed privacy-preserving data mining (PPDM). PPDM technique or approach enables the extraction of actionable insight from enormous volume of data while safeguarding the privacy of individual information and benefiting the entire society*

*Medical research has taken a new course as a result of data mining with healthcare data to detect diseases earlier and improve patient care. Data integration necessitates the sharing of sensitive patient information. However, substantial privacy issues are raised in connection with the storage and transmission of potentially sensitive information. Disclosing sensitive information infringes on patients' privacy. This paper aims to conduct a review of related work on privacy-preserving mechanisms, data protection regulations, and mitigating tactics. The review concluded that no single strategy outperforms all others. Hence, future research should focus on adequate techniques for privacy solutions in the age of massive medical data and the standardization of evaluation standards.*

**Keywords:** *Privacy-Preserving Data Mining. Algorithm. Electronic Health Records (EHR). Perturbation. Randomization. Additive Noise. Cryptography. Anonymization. Condensation.*

## 1.0 Introduction

In recent years, the use of computer science in everyday activities has increased at an unprecedented rate. Organizations, communities, and people all exhibit an increased trend toward electronic data storage. This development has influenced the healthcare industry into undergoing a data revolution that could lead to significant improvements in patient care, medical research, and public health (Lv & Qiao, 2020). However, this data revolution is taking place on a global scale. Data mining has become more and more common in the healthcare industry, which is been used to mine large datasets for insightful patterns and insights using cutting-edge analytical methods. Nevertheless, despite the benefits that data mining in healthcare can provide, it is urgently necessary to address the critical problem of maintaining the safety and confidentiality of sensitive patient data. This paper endeavors to provide an inclusive literature review centered on the techniques, classification and the scenarios of their implications on various techniques applied in preserving the safety and confidentiality of data mining information in health care.

## 1.1 Background and Context of Data Mining in Health Care

Data mining, also known as knowledge discovery in databases (KDD), refers to the process of obtaining important facts and arrangements from an immense quantity of data (Fayyad, 1997). In the perspective of healthcare, data mining encompasses the analysis of clinical probationary data, medical imaging data, electronic health records (EHRs), and other patient-related information to reveal hidden insights that can advance medical diagnosis, treatment efficiency, and patient results (Ojha, & Mathur, 2016).

Data mining, according to Fayyad et al. (1996), is the process of discovering new, meaningful correlations, patterns, and tendencies by sifting through enormous amounts of data that have been deposited in sources. This is done using configuration recognition technologies as well as statistical and mathematical techniques. These researchers emphasized the significance of employing a variety of investigative techniques to extract useful information from sizable datasets. Data mining, according to Han et al. (2011), it is the process of taking substantial amounts of data kept in databases, data warehouses, or other information repositories and extracting valuable information from it. The article emphasized that data mining aims to uncover significant patterns and insights that can be applied to decision-making and the acquisition of competitive advantages. Lee et al., (2016) emphasized the importance of discovering hidden patterns and facts that can be predicted, classification, and decision-making.

However, several scholars have extensively studied how data mining has been employed in the healthcare sector, uncovering valuable insights and patterns to improve patient care, medical research, and public health. The implementation of electronic health record structures and digital health technologies has rapt an exponential fruition in healthcare data, making data mining systems even more pertinent and impactful in this sector. Bellazzi and Zupan (2008) discussed data mining in health care as the method of extracting hidden designs from theoretically large datasets for medical verdict support. The scholars underscored the significance of data mining in assisting medical professionals with clinical decision-making, disease prediction, and treatment planning. Aslo, Malik et al., (2018) described data mining in health care as the use of advanced data analysis techniques to discover patterns and trends in health care data for improving medical decision-making and health outcomes.

Sophisticated pattern extraction has been made possible by the rise in capacity and inventions of the contemporary period and the collection and analysis of vast quantities of data. As a result, the political, industrial, and medical sectors now have access to invaluable knowledge that benefits governing societies and companies. (Desmet & Cook, 2021). Additionally, Kim et al. (2018) point out that interest in innovative services for healthcare, which may offer customers high-quality healthcare whenever and wherever they are, has increased due to the shift in healthcare services from the curative to the preventative approach. This led to the development of numerous devices. However, the proliferation of the internet of things, according to Chamikara et al. (2020), may lead to a breach of privacy during the data mining process used to produce actionable intelligence for decision making. The need to decontaminate personally identifiable information before allowing it to be used for investigation is becoming more and more important, despite the fact that sharing knowledge is valuable (Ehrmann & Stinson, 1999). It is possible to draw the conclusion that the development of actionable intelligence in the field of health care big data analysis has had a significant positive impact on society's economic and social well-being given the advance of big data, innovations in health care, and scientific methods that are continuously updated. Although the vast amounts of data produced by healthcare organizations can help doctors predict and diagnose diseases, worries about privacy and security have become more prevalent (Milovic & Milovic, 2012). These advantages have made it more challenging to prevent the safety and confidentiality risks connected to identifying and abusing mined data.

According to Domadiya & Rao (2021), the collection of healthcare data on an unpredictable central server poses a threat to privacy. In order to prevent patient apathy toward disclosing their data, information safety and confidentiality in health care data mining are crucial. Additionally, each collaborative participant will need to share personal data for data mining due to the distributed environment. Although these patterns add significant knowledge to a particular field, the sharing of sensitive data raises privacy issues. According to Sing (2020), the demand for better data safety and confidentiality is growing daily. According to Zainab and Kechadi (2019), privacy-preserving data mining is a method for drawing knowledge from data while upholding privacy.

This literature review summarizes some research on a technique for preserving sensitive information in health care when data mining through various PPDM algorithms and techniques.

## 2.0    LITERATURE REVIEW

This Chapter delve into the theoretical framework of data safety and confidentiality in the context of data mining in the healthcare industry. This chapter provides an in-depth exploration of the foundational concepts and principles that underpin data safety and confidentiality, focusing on their significance in safeguarding sensitive health data during data mining processes. Theoretical perceptions from intellectual exploration and literature are used to explicate the key ideas in this domain.

### 2.1    CONCEPTUAL REVIEW

#### 2.1.1    Data Security Measures

Data security procedures refer to a set of protective travels and protocols designed to safeguard complex data from unlicensed access, disclosure, alteration, or obliteration (Sripriyanka & Mahendran, 2022). These procedures aim to certify the confidentiality, availability of data, and integrity, principally in the framework of data mining in health care. According to Bartolacci and Antonelli, (2017), these procedures may include encryption, access controls, intrusion detection systems, firewalls, and regular data backings.

For Lupa and Gogan (2014), Data security processes encompass the application of technical, administrative, and physical controls to keep data integrity, confidentiality, and availability. These processes address probable risks and susceptibilities, ensuring that data remains safe throughout its lifecycle. Similarly, Omotunde & Ahmed (2023) opines that data security measures comprise an inclusive methodology to safeguarding data assets, including certification mechanisms, encryption, data access controls, data masking, and auditing. Protecting sensitive information from both internal and external threats is the goal of these procedures.

However, one indispensable aspect of data security measures includes technical safeguards that use technology to shield data. Encryption is a broadly adopted system to render data unreadable without the appropriate decryption key, guaranteeing that even if unauthorized personages gain access to the data, they cannot decipher it (Krisby, 2018). Additionally, access controls restrict data access to authorized users based on their roles, privileges, or authentication credentials, thereby preventing unauthorized users from accessing sensitive information. Data security measures extend beyond traditional safeguards and focus on the importance of data governance, data lifecycle management, and privacy-by-design ideologies. Emphasizing accountability and transparency in data handling is crucial for effective data security (Kshetri, 2014).

### 2.1.2　Data Privacy Regulations

Rubinstein (2012) refer to Data privacy regulations the legal frameworks and guidelines that govern the collection, use, storage, sharing, and protection of personal data. These regulations are designed to ensure that individuals' privacy rights are respected and that organizations handling personal data adhere to specific requirements and standards. These regulations according to Chen and Zhao (2012) outline the responsibilities and obligations of data controllers and processors, impose restrictions on data usage, and establish mechanisms for obtaining informed consent from data subjects. Compliance with data privacy regulations is essential to maintain the confidentiality, integrity, and availability of personal data while respecting individuals' rights to data protection and privacy. "These regulations aim to strike a balance between facilitating data-driven innovations and ensuring that individuals' privacy rights are respected. Organizations must comply with these regulations by implementing data protection measures, obtaining valid consent, and notifying data breaches promptly to protect data subjects' privacy and maintain public trust." (Tikkinen-Piri, et al., 2018).

### 2.1.3 Privacy-Preserving Technologies in Use

Data security has been the subject of numerous research studies, which have sparked the creation of numerous data mining techniques and algorithms. In order to protect or preserve the privacy of sensitive data, privacy preserving data mining (PPDM) is used. It is known as PPDM. In this paper, some PPDM techniques are reviewed, such as:

#### 2.1.3.1 Perturbation

As proposed by Zainab and Kechadi's (2019), sensitive values should be replaced with some false values during perturbation so that the statistical values, which are the generated data closely resemble the original data. However, Chamikara et al. (2020) propose a perturbation (modification) algorithm for Big Data that combines an innovative privacy model with an irreversible input perturbation mechanism known as optimal geometric transformations (PABIDOT) to enable complete data release while enhancing attack resistance, classification accuracy, and efficiency when working with large amounts of data. While Siraj et al. (2019) reported that though perturbation preserves statistical information and averts the attacker from recovering sensitive information from the data or execute sensitive linkage, human finds the perturbed record pointless. Desmet & Cook (2021) submit that increasing the amount of perturbation reduces the data set's utility.

#### 2.1.3.2 Condensation

In Siraj et al. (2019), the condensation method creates constrained clusters within a dataset and generates pseudo data that serves as an additional line of defense against adversarial attacks. However, there are some drawbacks, including information loss and the effort needed to compress a large dataset into a statistical group. Data condensation, according to Zainab & Kechadi (2019), produces better results but might change the format of the data.

#### 2.1.3.3 Anonymization

By removing explicit identifiers, Zainab & Kechadi (2019) showed that while using an anonymization approach ensures the security of individual private data, there is still a risk of privacy breaches known as linking attacks.

This happens when quasi-identifiers are combined with publicly available data, and Pika et al. (2020) reported that the impact of anonymization varies depending on the process mining algorithm and the log characteristics. According to Zainab & Kechadi (2019), it only applies to centralized data.

#### 2.1.3.4  Randomization

In the word of Zainab & Kechadi (2019), randomization is a low-cost and effective technique for PPDM, and the result is data on which nobody can be sure whether it is true or false.

However, Desmet & Cook (2021) explain that random methods can be highly effective; though, attempting to resolve this issue by the addition of noise may affect the data's integrity. As a result, randomization is insufficient for providing privacy for non-continuous data.

#### 2.1.3.5 Cryptography

Cryptography techniques can compromise privacy when many parties are involved, claim Zainab & Kechadi (2019). Singh (2020) suggests watermarking be added to current data concealing systems to increase information security. While providing a secure communication channel, cryptography also guards against the leakage of private data during resource sharing between multiple parties (Singh, 2020). According to Zainab & Kechadi (2019), this could put the results of data mining at risk when information is shared.  To manage and enhance healthcare big data safety and confidentiality, Sharma et al. (2021) suggest a Blockchain-based architecture called smart contract-based architecture. This architecture maintains data protection while reducing processing overhead and allowing users to confirm the accuracy of requested files (Li et al., 2013).

#### 2.1.3.6  Additive Noise

Desmet & Cook (2021) reported that, adding noise results in quantifiable uncertainty, demonstrating a clear trade-off between data use and privacy. Local differential privacy (LDP), a method introduced by Kim et al. (2018), is a method for gathering personal health information streams. The idea behind it is to ensure that the data contributor's data is protected throughout the data collection process by modifying the original data with carefully chosen random noise before transmitting it to a collector.

Although Siraj et al. (2019) reported that, despite its simplicity, the additive noise technique is no longer used to estimate the initial values of individual records, Pika et al. (2020) reported that noise addition preserves the aggregate distribution of attribute values.

**2.1.3.7    Algorithm**

Using the 3D rotation data perturbation (RDP) and Singular Value decomposition (SVD) machine learning algorithms, Kousika & Premalatha (2021) proposed a privacy-preserving methodology for data preservation. They reported that the proposed methodology outperforms rivals and strikes a balance between utility and the privacy of data while maintaining dataset confidentiality. However, Siraj et al. (2019) reported that association rules and clustering are used to prevent the disclosure of sensitive information, and performance, level of uncertainty, randomization, and utility of the data are used to evaluate the effectiveness of privacy-preserving data mining methods.

Wu et al. (2021) discussed the use of an algorithm called Ant Colony System to Data Mining (ACS2DTM), which employs constraints (multi threshold) to protect and sanitize the varying lengths of patient records while still maintaining valuable insight for mining purposes.

**2.2    EMPIRICAL REVIEW**

The health data comes from diverse sources. As a result, the safety and confidentiality of health personal identifiable information cannot be overstated due to malicious intent to profit from the data's use at the expense of the patient (Wasserman and Wasserman, 2022). An attacker can easily mine health data, particularly vulnerable, and release it to the public. Abouelmehdi et al., (2018) define privacy as the ability to safeguard information that can be used to identify a certain individual's health. It centers on the usage and administration of a person's personal information, such as formulating rules and implementing permission procedures to guarantee the appropriate collection, sharing, and use of patients' personal information.  Also, security is described as the prevention of unauthorized, including preserving integrity and availability (Sood, 2012). In line with the preceding, it is self-evident that security is sufficient to ensure the privacy of large amounts of health data.

Abouelmehdi et al. (2018) in the research work titled "Big healthcare data: Preserving safety and confidentiality" noted that big healthcare data has considerable potential to improve patient outcomes, predict outbreaks of epidemics, gain valuable insights, avoid preventable diseases, reduce the cost of healthcare delivery and improve the quality of life in general. However, determining on the permissible uses of data while conserving security and patient's right to privacy is a difficult task. According to the research, big data, no matter how useful for the advancement of medical science and vital to the success of all healthcare organizations, can only be used if safety and confidentiality issues are addressed. The paper surveyed the state-of-the-art safety and confidentiality challenges in big data as applied to healthcare industry, assessed how safety and confidentiality issues occur in case of big healthcare data and discussed ways in which they may be addressed. However, the study mainly focused on the anticipated methods based on anonymization and encryption, associated their assets and confines, and projected future research directions.

Thapa, & Camtepe, (2021) also highlight the importance of security in every phase of big data. In a research carried out on the requirements, challenges, and existing techniques for data safety and confidentiality. The study leverages information from various sources, including omics, lifestyle, environment, social media, medical records, and medical insurance claims to enable personalized care, prevent and predict illness, and precise treatments. The scholars noted that health data contain sensitive private information, which include patient identity, career and the patient medical situations, therefore, proper care is required at all times. However, leakage of this private information affects the personal life, including bullying, high insurance premium, and loss of job due to the health history. Thus, the security, privacy of and trust on the information are of utmost importance. Subsequently, as the public is the targeted beneficiary of the system, the effectiveness of precision health reduces. Herein, in the light of precision health data security, privacy, ethical and regulatory requirements, finding the best methods and techniques for the utilization of the health data, and thus precision health is essential.

In addition, Pika et al. (2020) in "Privacy-Preserving Process Mining in Healthcare" demonstrate that process mining has been successfully applied in the healthcare domain and has helped to uncover various insights for improving healthcare processes. The article analyse data privacy and utility requirements for healthcare process data and assess the suitability of privacy-preserving data transformation methods to anonymise healthcare data. The study demonstrate how some of these anonymisation methods affect various process mining results using three publicly available healthcare event logs. It describe a framework for privacy-preserving process mining that can support healthcare process mining analyses. It also advocate the recording of privacy metadata to capture information about privacy-preserving transformations performed on an event log.

According to some estimates from the Health Insurance Portability and Accountability Act (HIPAA), in 2020, there were more data intrusions in health care recorded, at a rate of more than 1.76 per day, and over 29 million healthcare records were compromised (Alder, 2021).

Desmet and Cook (2021) propose two classes of PPDM methods and mitigation strategies for overcoming the difficulties associated with mitigating attack vulnerabilities or with maintaining the location information privacy: a merger of PPDM procedures with those in the cryptography sector, as well as a blending of multiple PPDM approaches. Lv & Qiao (2020) examined a sample of questionnaires distributed to Chinese physicians, nurses, administrators, and technicians. The results indicated that when medical care big data is oriented toward cloud services, the probability of data analysis, medical treatment processes, disease diagnosis processes, a lack of protective measures and an imperfect access system are all-greater than 80%. Therefore, attention should be paid the access.

Based on the experiment, Lv & Qiao (2020) proposed technology and management as two levels of privacy protection measures, and concluded that medical institutions should prioritize data privacy protection, strengthen their protection measures to prevent patient privacy disclosure and malicious attacks on the medical system by external hackers. As well, grasp the use of digital medical data to provide decision support for subsequent medical data analysis.

Thapa et al. (2021) propose cryptographic security, blockchain-based security, access control, security analysis, network security, anonymization, and pseudonymization techniques for Precision health data at rest and in transit. Furthermore, techniques like; trusted execution environment, homomorphic encryption, multiparty computation, and differential privacy (DP) were proposed for data in use. Furthermore, Thapa & Camtepe (2021) assert that blockchain, cryptography, hardware-based techniques, differential privacy, and consent and privacy policies are the most widely available techniques for securing and protecting precision health data.

As a result of the preceding, no single technology can provide comprehensive safety and confidentiality solutions; multiple technologies are required. Lv & Qiao (2020) proposed that in the future, a risk prediction system for health care big data should be developed to bolster the security of health care big data.

## 2.3     THEORETICAL REVIEW

To achieve the aims of this research work, several theories can be used to provide a robust theoretical foundation and analytical framework. These theories shall assist in understanding the underlying principles, challenges, and implications related to data safety and confidentiality in the context of health care data mining. Therefore, the theories adopted for this research are the Privacy Calculus Theory and the Control Theory.

### 2.3.1     The Privacy Calculus Theory

Privacy Calculus Theory, also known as the Privacy Calculus Model, is a theoretical framework used to understand individuals' decision-making processes regarding privacy-related choices. The theory suggests that individuals engage in a cost-benefit analysis when making decisions about disclosing personal information or engaging in activities that may impact their privacy. Alan D. Westin, an American professor of public law, introduced the Privacy Calculus Theory and government, in his book titled "Privacy and Freedom," first published in 1967. In the book, Westin explored the concept of privacy and discussed how individuals make decisions regarding the disclosure of their personal information in various contexts.

The Privacy Calculus Theory however can be applied to analyze individuals' decisions regarding the disclosure of their health data for data mining in the healthcare industry. According to the theory, individuals make decisions about sharing personal information based on a cost-benefit analysis, where the perceived benefits of sharing the data outweigh the perceived risks to privacy. To examine how the Privacy Calculus Theory can be relevant to this literature review. Applying the Privacy Calculus Theory, we can understand that patients and individuals may weigh the benefits of improved healthcare outcomes and medical advancements against the potential risks of data breaches and privacy violations when deciding whether to share their health data for data mining purposes.

The scholars' concepts of data security measures and data privacy regulations are critical factors influencing the privacy calculus of individuals in the healthcare context. Data security measures involve technical, administrative, and physical controls designed to protect data integrity, confidentiality, and availability (Domadiya & Rao, 2021). On the other hand, data privacy regulations refer to legal frameworks that govern the collection, use, and protection of personal data, ensuring individuals' rights to data protection and privacy (Smith & Johnson, 2019).

In the literature review section, various privacy-preserving techniques for data mining in healthcare are discussed, such as perturbation, condensation, anonymization, randomization, cryptography, additive noise, and algorithm-based methods (Zainab & Kechadi, 2019; Desmet & Cook, 2021; Pika et al., 2020). These techniques denote the efforts to strike stability between data utility and privacy preservation, permitting researchers and healthcare professionals to improvement valuable insights while protecting patients' sensitive material. Applying the Privacy Calculus Theory, patients and individuals may measure the effectiveness of these privacy-preserving techniques in anonymizing their data and preventing latent privacy breaches. They may weigh the benefits of contributing to medical research and improving healthcare outcomes against the potential risks of re-identification and unauthorized access to their health data.

The Privacy Calculus Theory can be used to better understand how patients and people decide whether to share their health data for data mining. It sheds light on how patients balance their worries about data safety and confidentiality against the potential advantages of data-driven medical advancements. To build trust and encourage data sharing for significant medical advancements, healthcare organizations and researchers must comprehend the privacy calculus of individuals in order to design effective data security measures and comply with data privacy regulations.

### 2.3.2 Control Theory

A psychological and sociological theory called "control theory" looks to control, regulation, and feedback as mechanisms for understanding human behavior. It is predicated on the idea that people are driven to preserve a sense of control over their lives and environments and that they engage in behaviors and actions to accomplish this sense of control. In the late 1960s and early 1970s, American psychiatrist and psychologist William Glasser proposed the control Theory. As part of his larger contributions to the fields of psychiatry and psychology, Glasser created the Control Theory.

However, Kitchin (2014) notes that the growing application of computer science in daily life has sparked a data revolution in the healthcare sector. Data mining, also referred to as knowledge discovery in databases (KDD), is important for sifting through mammoth amounts of healthcare data, such as electronic health records, medical imaging data, and clinical trial data, to find valuable patterns and insights (Bibri, 2018). Medical research, patient care, and public health could all be significantly improved by the insights gleaned from data mining.

Nevertheless, the safety and privacy of sensitive patient data have come under study due to this data revolution. According to Kimmons and Veletsianos (2018), large datasets are frequently analyzed using data mining techniques, and the conclusions drawn may include personally identifiable information. To avoid potential breaches, unauthorized access, or misuse of sensitive data, maintaining the safety and confidentiality of this data is crucial.

Control Theory, as applied to this context, emphasizes the need for organizations and healthcare providers to have control over their data and the decision-making processes involved in data mining. Data security measures are essential protective actions and protocols that organizations should implement to safeguard sensitive data from unauthorized access, disclosure, alteration, or destruction (Peltier, 2016). These measures may include encryption, access controls, intrusion detection systems, firewalls, and regular data backups. By implementing technical, administrative, and physical controls, organizations can ensure the confidentiality, integrity, and availability of data (Cheung, 2014).

Data privacy regulations are legal frameworks that govern the collection, use, storage, sharing, and protection of personal data. These regulations aim to strike a balance between promoting data-driven innovations and respecting individuals' rights to privacy (Zyskind & Nathan, 2015). Organizations must comply with these regulations by implementing data protection measures, obtaining valid consent, and promptly reporting data breaches. To preserve privacy while conducting data mining in healthcare, several privacy-preserving techniques are available. These techniques aim to extract meaningful insights from data without compromising individual privacy. Some of the techniques include perturbation, condensation, anonymization, randomization, cryptography, additive noise, and algorithm-based privacy-preserving methodologies. Each technique has its advantages and limitations, and organizations must carefully choose the most appropriate method based on their specific requirements.

Despite the benefits of data mining in healthcare, concerns about privacy and security persist. Privacy-preserving data mining approaches are vital to ensure that sensitive information remains protected during the data mining process (Qi, & Zong, 2012). This involves striking a balance between data utility and privacy, implementing robust security measures, and adhering to relevant data privacy regulations.

## 3.0 FINDINGS

This literature review shows that advances in data mining have spawned a new field of a study dubbed privacy-preserving data mining. PPDM technology enables extraction of actionable intelligence from large data set while safeguarding the privacy of individual records.

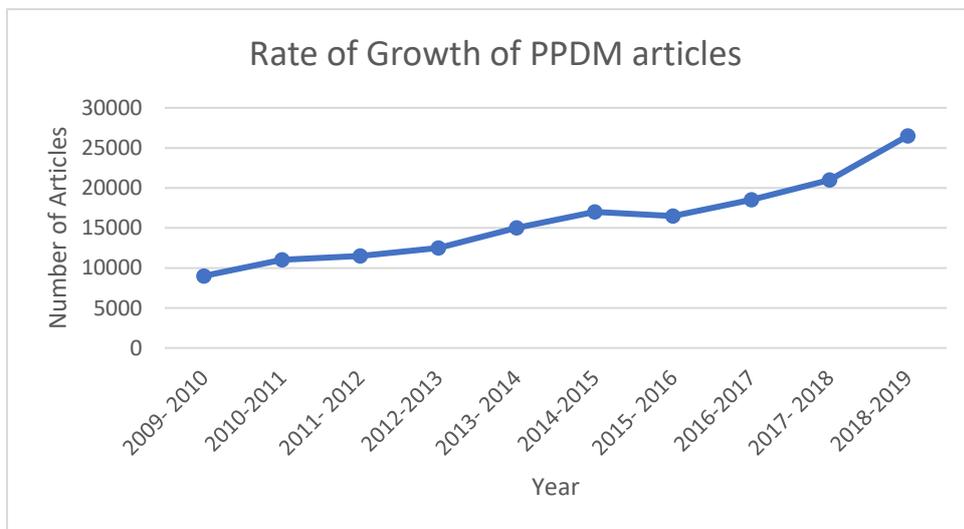

**Adapted source:** Desmet & Cook (2021). The growth rate of PPDM Articles

Privacy-preserving data mining demonstrates the ability to use data mining tools without revealing confidential information. In other words, the proposed sed methodology ensures the confidentiality of data and the secure sharing of data. However, each of these techniques and approaches has demonstrated some limitations. It is deducible. None of the methods alone can provide a complete safety and confidentiality solution; multiple techniques must be used in combination. Greater emphasis should be placed on the security of potentially personally identifiable information.

The year 2020 saw an exponential increase in health care breaches, indicating that the issue of healthcare safety and confidentiality has received little attention (Alder, 2021). Additionally, the data protection law must consider factors other than fundamental human rights when ensuring the safety and confidentiality of data. Wu et al. (2021) reported that safety and confidentiality are frequently sacrificed for usability in an era of global pandemics.

## 4.0 CONCLUSION

In conclusion, this literature review highlights the significant impact of data mining in the healthcare industry and the potential it holds for improving patient care, medical research, and public health. As the use of computer science and digital technologies continues to grow, healthcare organizations are experiencing a data revolution with the increasing adoption of electronic health records and other patient-related information.

However, amidst the benefits, the review also emphasizes the critical importance of addressing the safety and confidentiality challenges associated with data mining in healthcare. Patient data is sensitive and personal, and any breach or misuse can have severe consequences for individuals and erode public trust in the healthcare system. Privacy-preserving data mining techniques have been developed to strike a balance between data utility and individual privacy, ensuring that valuable insights can be derived without compromising sensitive information.

The Privacy Calculus Theory and Control Theory provide theoretical frameworks to understand patients and individuals' decision-making processes regarding data sharing and privacy concerns. Patients weigh the potential benefits of data-driven medical advancements against their concerns about data safety and confidentiality before consenting to share their health data.

It is evident from the literature review that data safety and confidentiality in healthcare are complex issues that require continuous attention and proactive measures. As data mining technology continues to evolve, healthcare organizations must keep up with the latest advancements in privacy-preserving techniques and adhere to data privacy regulations to maintain patient trust and confidence in the healthcare system.

In conclusion, while data mining has tremendous potential for transformative advancements in healthcare, ensuring data safety and confidentiality must be a top priority for healthcare providers and researchers. Striking the right balance between data utility and individual privacy will lead to a safer and more effective data-driven healthcare landscape. Continued research and advancements in privacy-preserving data mining techniques will be crucial to meet the growing demands of data analysis while safeguarding patient information. Ultimately, by addressing the safety and confidentiality challenges in healthcare data mining, we can unlock the full potential of data-driven healthcare for the benefit of patients, medical professionals, and society as a whole.

**REFERENCES**

Abouelmehdi, K., Beni-Hessane, A. & Khaloufi, Hayat. (2018). Big healthcare data: Preserving Security and Privacy. *Journal of Big Data, 5(1),* 1–18. https://doi.org/10.1186/s40537-017-0110-7

Alder, Steve (2021, January 19). 2020 Healthcare data breach report: 25% increase in breaches in 2020. HIPAA Journal. https://www.hipaajournal.com/2020-healthcare- data-breach-report-us/

Bellazzi, R., & Zupan, B. (2008). Predictive data mining in clinical medicine: current issues and guidelines. *International journal of medical informatics*, *77*(2), 81-97.

Bibri, S. E. (2018). The IoT for smart sustainable cities of the future: An analytical framework for sensor-based big data applications for environmental sustainability. *Sustainable cities and society*, *38*, 230-253.

Chamikara, M. A. P., Bertok, P., Liu, D., Camtepe, S., & Khalil, I. (2020). Efficient privacy preservation of big data for accurate data mining. *Information Sciences, 527,* 420–443. https://doi.org/10.1016/j.ins.2019.05.053

Chen, D., & Zhao, H. (2012, March). Data security and privacy protection issues in cloud computing. In *2012 international conference on computer science and electronics engineering* (Vol. 1, pp. 647-651). IEEE.

Cheung, S. K. (2014). Information security management for higher education institutions. In *Intelligent Data analysis and its Applications, Volume I: Proceeding of the First Euro-China Conference on Intelligent Data Analysis and Applications, June 13-15, 2014, Shenzhen, China* (pp. 11-19). Springer International Publishing.

Comandè, G., & Schneider, G. (2018). Regulatory challenges of data mining practices: The case of the Never-ending Lifecycles of 'Health Data.' *European Journal of Health Law, 25(3)*, 284–307. https://doi-org.ezproxy.nwmissouri.edu/10.1163/15718093-12520368

Desmet, C., & Cook, D. J. (2021). Recent developments in privacy-preserving mining of clinical data. *ACM/IMS Transactions on Data Science, 2(4).* https://doi.org/10.1145/3447774

Domadiya, N., & Rao, U. P. (2019). Privacy preserving distributed association rule mining approach on vertically partitioned healthcare data. *Procedia Computer Science, 148*, 303–312. https://doi-org.ezproxy.nwmissouri.edu/10.1016/j.procs.2019.01.023

Domadiya, N., & Rao, U. P. (2021). Improving healthcare services using source anonymous scheme with privacy preserving distributed healthcare data collection and mining. *Computing, 103(1)*, 155–177. https://doi-org.ezproxy.nwmissouri.edu/10.1007/s00607-020-00847-0

Ehrmann, J. R., & Stinson, B. L. (1999). JOINT FACT-FINDING AND THE. *The consensus building handbook: A comprehensive guide to reaching agreement*, 375.

Fayyad, U. (1997, August). Data mining and knowledge discovery in databases: implications for scientific databases. In *Proceedings. Ninth International Conference on Scientific and Statistical Database Management (Cat. No. 97TB100150)* (pp. 2-11). IEEE.

Han, J., Kamber, M., & Pei, J. (2011). Data mining: concepts and techniques. Morgan Kaufmann.

Kim, J. W., Jang, B., & Yoo, H. (2018). Privacy-preserving aggregation of personal health data streams. *PLoS ONE, 13(11)*, 1–15. https://doi-org.ezproxy.nwmissouri.edu/10.1371/journal.pone.0207639

Kimmons, R., & Veletsianos, G. (2018). Public internet data mining methods in instructional design, educational technology, and online learning research. *TechTrends*, *62*(5), 492-500.

Kitchin, R. (2014). The data revolution: Big data, open data, data infrastructures and their consequences. *The Data Revolution*, 1-240.

Kousika, N., & Premalatha, K. (2021). An improved privacy-preserving data mining technique using singular value decomposition with three-dimensional rotation data perturbation. *Journal of Supercomputing, 77(9)*, 10003–10011. https://doi-org.ezproxy.nwmissouri.edu/10.1007/s11227-021-03643-5

Krisby, R. M. (2018). Health care held ransom: Modifications to data breach security & the future of health care privacy protection. *Health Matrix*, *28*, 365.

Kshetri, N. (2014). Big data׳s impact on privacy, security and consumer welfare. *Telecommunications Policy*, *38*(11), 1134-1145.

Lee, M., Kwon, W., & Back, K. J. (2021). Artificial intelligence for hospitality big data analytics: developing a prediction model of restaurant review helpfulness for customer decision-making. *International Journal of Contemporary Hospitality Management*, *33*(6), 2117-2136.

Lv, Zhihan, & Qiao, L. (2020). Analysis of healthcare big data. *Future Generation Computer Systems, 109*, 103–110. https://doi- org.ezproxy.nwmissouri.edu/10.1016/j.future.2020.03.039

Malik, M. M., Abdallah, S., & Ala'raj, M. (2018). Data mining and predictive analytics applications for the delivery of healthcare services: a systematic literature review. *Annals of Operations Research*, *270*, 287-312.

Milovic, B., & Milovic, M. (2012). Prediction and decision making in health care using data mining. *Kuwait chapter of arabian journal of business and management review*, *1*(12), 1-11.

Ojha, M., & Mathur, K. (2016, March). Proposed application of big data analytics in healthcare at Maharaja Yeshwantrao Hospital. In *2016 3rd MEC International Conference on Big Data and Smart City (ICBDSC)* (pp. 1-7). IEEE.

Omotunde, H., & Ahmed, M. (2023). A Comprehensive Review of Security Measures in Database Systems: Assessing Authentication, Access Control, and Beyond. *Mesopotamian Journal of CyberSecurity*, *2023*, 115-133.

Peltier, T. R. (2016). *Information Security Policies, Procedures, and Standards: guidelines for effective information security management*. CRC press.

Pika, Anastasiia, Wynn, Moe T., Budiono, Stephanus, Hofstede, Arthur H.M. ter, Van der Aalst, Wil M.P., & Reijers, Hajo A. (2020). Privacy-preserving process mining in healthcare. *International Journal of Environmental Research and Public Health, 17(5)*, 1612. https://doi-org.ezproxy.nwmissouri.edu/10.3390/ijerph17051612

Qi, X., & Zong, M. (2012). An overview of privacy preserving data mining. *Procedia Environmental Sciences*, *12*, 1341-1347.

Sharma, P., Borah, M. D., & Namasudra, S. (2021). Improving security of medical big data by using Blockchain technology. *Computers and Electrical Engineering, 96(Part A)*. https://doi-org.ezproxy.nwmissouri.edu/10.1016/j.compeleceng.2021.107529

Singh, A. K. (2020). Data hiding: Current trends, innovation and potential challenges. *ACM Transactions on Multimedia Computing, Communications & Applications, 16(3s),* 1–16

Siraj, M. M., Rahmat, N. A., & Din, M. M. (2019). A survey on privacy-preserving data mining approaches and techniques. *Proceedings of the 2019 8th International Conference on Software and Computer Applications,* 65-69. https://doi.org/10.1145/3316615.3316632

Sood, S. K. (2012). A combined approach to ensure data security in cloud computing. *Journal of Network and Computer Applications*, *35*(6), 1831-1838.

Sripriyanka, G., & Mahendran, A. (2022). Bio-inspired Computing Techniques for Data Security Challenges and Controls. *SN Computer Science*, *3*(6), 427.

Thapa, C., & Camtepe, S. (2021). Precision health data: Requirements, challenges and existing techniques for data security and privacy. *Computers in Biology and Medicine, 129*. https://doi-org.ezproxy.nwmissouri.edu/10.1016/j.compbiomed.2020.104130

Tikkinen-Piri, C., Rohunen, A., & Markkula, J. (2018). EU General Data Protection Regulation: Changes and implications for personal data collecting companies. *Computer Law & Security Review*, *34*(1), 134-153.

Wu, J. M.-T., Srivastava, G., Lin, J. C.-W., & Teng, Q. (2021). A multi-threshold ant colony system-based sanitization model in shared medical environments. *ACM Transactions on Internet Technology, 21(2),* 1–26. https://doi.org/10.1145/3408296

Yadav, P., Steinbach, M., Kumar, V., & Simon, G. (2017). Mining electronic health records (EHRs): A survey. *ACM Computing Surveys, 50(6),* 1–40. https://doi.org/10.1145/3127881

Zainab, S. S., & Kechadi, T. (2019). Sensitive and private data analysis: A systemic review. *Proceedings of the 3rd International Conference on Future Networks and Distributed Systems,* 1-11. https://doi.org/10.1145/3341325.3342002

Zyskind, G., & Nathan, O. (2015, May). Decentralizing privacy: Using blockchain to protect personal data. In *2015 IEEE security and privacy workshops* (pp. 180-184). IEEE.